Liying Chen, Alexander M. McKillop, Ashley P. Fidler, and Marissa L. Weichman*

# Ultrafast optical modulation of vibrational strong coupling in ReCl(CO)$_3$(2,2-bipyridine)

**Abstract:** Polaritons – hybrid light-matter states formed from the strong coupling of a bright molecular transition with a confined photonic mode – may offer new opportunities for optical control of molecular behavior. Vibrational strong coupling (VSC) has been reported to impact ground-state chemical reactivity, but its influence on electronic excited-state dynamics remains unexplored. Here, we take a first step towards excited-state VSC by demonstrating optical modulation of the ReCl(CO)$_3$(bpy), (bpy = 2,2-bipyridine) complex under VSC using femtosecond ultraviolet (UV)-pump/infrared (IR)-probe spectroscopy. We establish ground-state VSC of ReCl(CO)$_3$(bpy) in a microfluidic Fabry-Pérot cavity equipped with indium tin oxide (ITO)-coated mirrors. ITO is effectively dichroic as it is reflective in the IR and transmissive in the UV-visible and therefore minimizes optical interference. Excitation with UV pump light drives ReCl(CO)$_3$(bpy) into a manifold of electronic excited states that subsequently undergo non-radiative relaxation dynamics. We probe the transient response of the strongly-coupled system in the mid-IR, observing both Rabi contraction and cavity-filtered excited state absorption signatures. We reconstruct the intrinsic response of intracavity molecules from the transient cavity transmission spectra to enable quantitative comparison with extracavity control experiments. We report no changes in the excited-state dynamics of ReCl(CO)$_3$(bpy) under ground-state VSC. However, we do observe significant amplification of transient vibrational signals due to classical cavity-enhanced optical effects. This effort lays the groundwork to pursue direct excited-state VSC aimed at modulating photochemical reactivity.

## 1 Introduction

The strong interaction of molecular vibrations with confined electromagnetic fields to form hybrid light-matter states known as polaritons has emerged as a potential new route for photonic control of molecular behavior [1], [2], [3], [4], [5], [6], [7], [8]. The emergence of vibrational polaritons appears to correlate with altered ground-state reactivity and vibrational energy redistribution [9], [10], [11], [12], [13], though many open questions remain surrounding the mechanisms and reproducibility of these effects [4], [14], [15], [16]. In any event, the early work in this field has stimulated broad interest in understanding how and when vibrational strong coupling (VSC) modulates molecular behavior. In this work, we take a first step towards examining how VSC may impact electronic excited state trajectories by optically exciting the ReCl(CO)$_3$(bpy), (bpy = 2,2-bipyridine) metal carbonyl complex under VSC in a microfluidic Fabry-Pérot (FP) optical cavity (Fig. 1AB).

Transient vibrational motions are sensitive reporters of excited state dynamics [17], [18] and can even influence photochemical outcomes [19], [20], [21]. Mode-selective infrared (IR) excitation of excited-state vibrations has accordingly been used to modulate charge transfer in donor-bridge-acceptor complexes [22], [23], [24], [25]. Such mode-selective excitation strategies face well-known challenges, however: competition with intramolecular vibrational relaxation and energy dissipation limits the scope of control over reaction outcomes. Achieving broadly-applicable vibrational modulation of excited-state dynamics remains an open research goal, motivating the search for new approaches, including cavity coupling. A particularly interesting possibility is the idea of engineering VSC in electronic excited states, where a cavity mode couples resonantly to a vibrational transition in the electronically excited manifold. While excited-state VSC has been proposed theoretically [26] and excited-state processes have been studied under ground-state VSC [27], actual experimental implementation of excited-state VSC remains elusive.

Achieving excited-state VSC requires a molecular candidate with: (a) a sufficiently strong IR-active vibrational mode to achieve VSC in the electronic excited state; (b) an optically bright electronic transition which can be pumped to populate the excited state; (c) distinct vibrational frequencies in the electronic excited and ground states to ensure resonant cavity-coupling only upon electronic excitation; and (d) sufficient population in the excited state to achieve collective excited-state VSC, even transiently. This last criterion is challenging to establish experimentally. Reaching the collective strong coupling regime relies on achieving a sufficiently large Rabi splitting, $\hbar\Omega_R$, which scales with $N^{1/2}$, where $N$ is the number of coupled molecules [2], [8].

Here, we identify the ReCl(CO)$_3$(bpy) complex as a target molecule that appears to meet criteria (a)–(c) above. ReCl(CO)$_3$(bpy) exhibits strong and well-characterized absorption features in both the mid-IR and the UV-visible (Fig. 1CD). The lowest electronic band lies between 350–400 nm (Fig. 1D) and corresponds to excitation from the $S_0$ electronic ground state to the manifold of singlet metal-to-ligand charge transfer ($^1$MLCT) excited states [28], [29]. As discussed by Šrut et al. [28], the $^1$MLCT band is dominated by the $S_1$, $S_2$, and $S_3$ electronic states, with $S_2$ expected to be the brightest state that will be initially populated by optical pumping. Excitation into the $^1$MLCT manifold leads to rapid intersystem crossing (ISC) into the triplet $^3$MLCT manifold on a 1–2 ps timescale, followed by electronic relaxation, vibrational cooling, and solvent reorganization on 5–10 ps timescales (Fig. 1A). Prior UV-pump/IR-probe experiments performed on ReCl(CO)$_3$(bpy) report significant blue-shifting of its three carbonyl stretching modes upon optical excitation [28], [30], [31], [32], [33]. For instance, the strongest-absorbing $a'$ carbonyl symmetric stretch lies at 2018 cm$^{-1}$ in the $S_0$ ground state (Fig. 1C) and appears at >2060 cm$^{-1}$ in the excited-state manifold [28], [30], [31], [32], [33]. The intensity of this symmetric carbonyl stretching mode, its large frequency shift upon photoexcitation, and its relative lack of spectral congestion make it a compelling target for modulation of ground-state VSC with UV excitation, as we demonstrate here, and perhaps eventually excited-state VSC.

*Corresponding author: Marissa L. Weichman, Department of Chemistry, Princeton University, Princeton, NJ, USA. E–mail: weichman@princeton.edu; https://orcid.org/0000–0002–2551–9146
**Liying Chen, Alexander M. McKillop, and Ashley P. Fidler**, Department of Chemistry, Princeton University, Princeton, NJ, USA.
**Ashley P. Fidler**, Current address: Chemistry Division, Naval Research Laboratory, Washington, DC 20375, USA.

Here, we perform ultrafast UV-pump/IR-probe spectroscopy on ReCl(CO)$_3$(bpy) in dimethyl sulfoxide (DMSO) solution embedded in a microfluidic FP optical cavity. We demonstrate VSC of the *a'* symmetric carbonyl stretching mode of ReCl(CO)$_3$(bpy) in the S$_0$ electronic ground state and modulate the collective coupling strength via optical pumping of population into the excited state manifold, as shown schematically in Fig. 1B. This result represents both a first step towards excited-state VSC and a new platform for cavity-mediated nonlinear optics. This system also provides a means to test whether ground-state VSC has any influence on excited-state dynamics. We ultimately find no statistically-significant change in the excited-state dynamics of ReCl(CO)$_3$(bpy) under ground-state VSC, as is perhaps expected given that the excited-state vibrations are both weakly and off-resonantly cavity-coupled in the current experiments.

Three important technical considerations underpin this work. First, we use indium tin oxide (ITO)-coated optics as dichroic cavity mirrors to enable UV-pump/IR-probe experiments with minimal optical artifacts. While the dichroic properties of ITO are well-known [34], its use as a mirror coating is not yet widespread for VSC. Secondly, we introduce a spectral reconstruction algorithm that allows us to extract the dynamics of strongly-coupled intracavity molecules, yielding observables that can be compared directly against extracavity controls. Transient spectroscopy of polaritonic systems can be a minefield of optical artifacts and interference effects that must be properly accounted for during analysis [35], [36], [37]. Our approach here is to reconstruct the intrinsic response of intracavity molecules by fitting transient cavity data to the analytical classical optics expression for transmission through an FP cavity. Finally, our measurements reveal a pronounced cavity-mediated enhancement of our transient pump-probe signals. This finding can be attributed to classical cavity-enhancement effects, e.g., the extreme sensitivity of the FP cavity transmission spectrum to minute changes in intracavity absorption.

Our results demonstrate the feasibility of optical modulation of VSC, tracking excited-state vibrational dynamics inside microfluidic cavity structures, and quantitative comparison of molecular dynamics across extracavity and intracavity experiments. We also provide a foundation for future efforts to achieve excited-state VSC, which would enable tests of cavity-modification of energy dissipation, charge transport, and photochemistry on excited state surfaces.

## 2 Methods and Materials

The ultrafast UV-pump/IR-probe spectrometer used herein represents a minor modification to the apparatus we have described previously [38], [39]. The system is based on a Ti:sapphire amplifier (Astrella, Coherent) which delivers 7 mJ, 60 fs pulses of 800 nm light at a 1 kHz repetition rate. UV pump pulses are generated using a tunable optical parametric amplifier (OPA, OPerA Solo with FH/SHSF options, Light Conversion), pumped with 2.5 mJ of the 800 nm fundamental to produce 46 µJ per pulse at 390 nm. Mid-IR probe pulses are produced using a second OPA (OPerA Solo with NDFG option, Light Conversion), pumped with 2.5 mJ of 800 nm fundamental light to produce 30 µJ pulses centered at 2040 cm$^{-1}$ (4900 nm) with a full-width at half-maximum (fwhm) bandwidth of 250 cm$^{-1}$.

We delay the mid-IR probe relative to the UV pump by up to 2.2 ns using a motorized stage (DL325, Newport). The pump and probe beams are spatially overlapped at the sample, with typical beam diameters of 200 µm. The probe strikes the sample at normal incidence while the pump is incident at a crossing angle of 7°. All measurements reported herein employ a magic-angle polarization scheme between pump and probe. Mid-IR probe light transmitted through the sample is spectrally dispersed using a diffraction grating and collected on a HgCdTe array detector (2DMCT, PhaseTech) operating in shot-to-shot mode, with the pump beam modulated at 500 Hz using a mechanical chopper (MC2000, Thorlabs). We calculate differential signals as $\Delta\text{OD} = -\log_{10}[\,I_{T,t}\,/\,I_{T,0}\,]$, where $I_{T,t}$ represents the intensity of probe light transmitted by the sample at time *t* after the pump pulse is incident and $I_{T,0}$ represents the "pump off" probe transmission spectrum acquired when the pump beam is blocked. We establish an instrument response function of 120 fs, comparable to that reported in our prior work [38]. Data is recorded using home-built LabVIEW programs and processed in MATLAB.

We prepare saturated (>40mM) solutions of ReCl(CO)$_3$(bpy) (99%, Strem Chemicals) in DMSO (≥99.9%, Sigma-Aldrich). We choose DMSO as solvent because it features high solubility for ReCl(CO)$_3$(bpy) and has been used in prior ultrafast studies of this complex, facilitating direct comparison with established literature [28], [31]. ReCl(CO)$_3$(bpy):DMSO solutions are flowed through a demountable microfluidic cell (TFC-M13-3, Harrick Scientific) using a peristaltic pump (Masterflex) to ensure continuous sample refreshment. For extracavity measurements, we assemble the flow cell with CaF$_2$ windows spaced with a 25 µm polytetrafluoroethylene (PTFE) spacer. For intracavity experiments, we fit the same flow cell with ITO-coated CaF$_2$ substrates (Colorado Concept Coatings) that serve as dichroic cavity mirrors. The ITO mirrors consist of a 20–30 nm SiO$_2$ base layer to adhere the coating to the substrate, a 150 nm (15 Ω/sq.) ITO layer, and a 50 nm SiO$_2$ protective overcoat. The microfluidic cell assembly and ITO mirrors are illustrated in Fig. 2A.

We use a commercial Fourier transform spectrometer (Nicolet iS50 FT-IR, Thermo Scientific) and UV-visible spectrometer (Cary 60 UV-Vis Spectrophotometer, Agilent) to characterize the broadband absorption spectra of extracavity solutions (Fig. 1CD), the reflectivity and transmission spectra of the ITO mirrors (Fig. 2BC), and the transmission spectra of empty FP cavities constructed from these ITO mirrors (Fig. 3A). All other transmission spectra of strongly-coupled cavity devices that we show herein are acquired using our ultrafast mid-IR beamline; these spectra are comparatively narrowband but the small spot size afforded by the collimated laser beam minimizes the impacts of cavity non-uniformity on transmission measurements [16]. We take advantage of deviations from perfect mirror parallelism to access distinct cavity detuning conditions by translating the cavity in the plane of the mirrors.

## 3 Results and Discussion

We first characterize the static optical properties of our dichroic ITO cavities and confirm that we can achieve VSC in solution-phase ReCl(CO)$_3$(bpy) samples in Section 3.1. We then review the ultrafast excited-state dynamics measured for extracavity ReCl(CO)$_3$(bpy) with UV-pump/IR-probe spectroscopy in Section 3.2. In Section 3.3, we detail UV-pump/IR-probe cavity transmission measurements performed for ReCl(CO)$_3$(bpy) under VSC, highlighting the distinct observables between intracavity measurements and extracavity controls. Finally, in Section 3.4 we introduce and apply a spectral reconstruction algorithm to recover the response of intracavity molecules independent of cavity filtering and directly compare the extracted intracavity dynamics against extracavity benchmarks.

### 3.1 Vibrational strong coupling of ReCl(CO)$_3$(bpy) in ITO cavities

We begin by characterizing the optical properties of the dichroic ITO mirrors and the resulting FP microcavities constructed from these mirrors. ITO is chosen as a mirror coating to provide sufficient reflectivity to achieve VSC in the mid-IR, while maintaining high transparency in the UV for efficient photoexcitation of the intracavity sample. Figure 2B plots the reflection and transmission spectra of a single ITO mirror across the mid-IR. The mirror reflectivity exceeds 90% in the carbonyl stretching region near 2000 cm$^{-1}$, which is of interest for VSC of ReCl(CO)$_3$(bpy). Meanwhile, Fig. 2C plots the transmission spectrum of a single ITO mirror across the UV-visible. The transmission exceeds 80% in the 350–400 nm region where the excited-state $^1$MLCT band of ReCl(CO)$_3$(bpy) lies. The transmission spectrum of a pair of ITO mirrors assembled into an empty 25 μm FP cavity is shown in Fig. 3A. The ITO cavities used herein feature typical free spectral ranges of 150 cm$^{-1}$ and empty-cavity mode linewidths of 15–18 cm$^{-1}$ fwhm.

We next confirm that ReCl(CO)$_3$(bpy) can be vibrationally strongly-coupled in such a cavity. To achieve resonant VSC, ReCl(CO)$_3$(bpy):DMSO solution is injected into a 25 μm FP cavity and a cavity mode is tuned into resonance with the desired molecular band by compressing the microfluidic clamp assembly to slightly change the cavity length. We couple to longitudinal FP cavity mode order $m \sim 13$–15. Figure 3B shows the static transmission spectrum of a representative cavity strongly-coupled to the $a'$ carbonyl symmetric stretching band of ReCl(CO)$_3$(bpy) at 2018 cm$^{-1}$ (solid light red trace) plotted against the free-space ReCl(CO)$_3$(bpy) absorption spectrum (blue trace). A simulated cavity transmission spectrum produced using the classical FP cavity expression discussed below in Section 3.4 is also plotted in Fig. 3B (red dotted trace). We observe a clear splitting of the cavity transmission peak into upper and lower vibrational polaritons (UP, LP). This system achieves a typical Rabi splitting of 24–30 cm$^{-1}$ which exceeds the linewidths of both the bare cavity mode (~15 cm$^{-1}$ fwhm) and molecular absorption line (~10 cm$^{-1}$ fwhm).

### 3.2 Extracavity excited-state vibrational dynamics

We now report on the ultrafast UV-pump/IR-probe spectroscopy of extracavity ReCl(CO)$_3$(bpy) to establish a baseline for its excited-state vibrational dynamics and benchmark against prior literature. We excite the system at 390 nm and record transient signals in the carbonyl stretching region near 2000 cm$^{-1}$ (Fig. 1A). Representative extracavity experimental data are presented in the left-hand column of Fig. 4. The free-space static absorption spectrum of ReCl(CO)$_3$(bpy) is reproduced in Fig. 4A, zoomed in on the $a'$ symmetric carbonyl stretch at 2018 cm$^{-1}$. Transient UV-pump/IR-probe spectra are presented in Fig. 4D as a function of wavenumber and pump-probe delay. Spectral lineouts at selected time delays are plotted in Fig. 4G.

The transient UV-pump/IR-probe spectra are dominated by a bleach feature centered near 2018 cm$^{-1}$ that appears within 1.5 ps (Fig. 4DG). We attribute this feature to the ground state bleach (GSB) of the $a'$ carbonyl stretching mode in the S$_0$ electronic ground state as population is transferred to the $^1$MLCT manifold [27], [28], [29], [30], [31], [32]. Concurrently, an excited-state absorption (ESA) band appears, initially centered at 2040 cm$^{-1}$ and blue-shifting past 2060 cm$^{-1}$ within 5–10 ps. This ESA feature is associated with the symmetric carbonyl stretching mode in the electronic excited states. The spectral shifting of the ESA feature is consistent with structural and electronic relaxation and solvent reorganization in the manifold of $^3$MLCT states [27], [28], [29], [30], [31], [32]. Both GSB and ESA signals decay gradually for time delays >100 ps, consistent with nonradiative relaxation and repopulation of the S$_0$ electronic ground state.

To quantify the observed extracavity dynamics, we take temporal lineouts at 2018 cm$^{-1}$ for the GSB and 2060 cm$^{-1}$ for the ESA. We then perform exponential fits to extract the short-term rise and long-term decay time constants for the GSB feature and the long-time decay constant for the ESA feature. To capture the frequency shifting of the ESA feature, we fit the ESA peak center for traces with time delay <50 ps using a Lorentzian and perform an exponential fit to the peak center as a function of time. More details and representative fits are presented in Section S1 of the Supplementary Material. Fitted extracavity time constants from individual experiments are laid out in Table S1, and values averaged across all experiments are summarized in Table 1. All extracavity observations are in agreement with prior studies of the excited state vibrational dynamics of ReCl(CO)$_3$(bpy) [27], [28], [29], [30], [31], [32] and provide a baseline for comparison with intracavity behavior.

### 3.3 Transient cavity transmission spectra

We now present UV-pump/IR-probe measurements of ReCl(CO)$_3$(bpy) under VSC. Representative experimental data are presented in the central column of Fig. 4. ReCl(CO)$_3$(bpy) is prepared under resonant VSC of its $a'$ symmetric carbonyl stretch at 2018 cm$^{-1}$, as described in Section 3.1. Fig. 4B reproduces the static strongly-coupled cavity transmission spectrum from Fig. 3B, acquired without any UV excitation. We excite the system at 390 nm through the high-transmission region of the ITO cavity mirrors and record transient cavity transmission signals in the carbonyl stretching region near 2000 cm$^{-1}$ (Fig. 1B). Transient UV-pump/IR-probe cavity transmission spectra are presented in Fig. 4E as a function of wavenumber and pump-probe delay, and spectral lineouts at selected time delays are plotted in Fig. 4H.

The raw transient cavity transmission spectra in the central column of Fig. 4 are, at first glance, markedly different than the extracavity transient absorption spectra in the left-hand column. Transient pump-induced changes in the intracavity complex refractive index alter the interference conditions that govern where cavity modes will appear and how much light they transmit [37]. Instead of distinct GSB and ESA peaks, the cavity transmission signals are dominated by derivative-like lineshapes centered at the polariton frequencies whose amplitudes and positions evolve as the system relaxes. These derivative lineshapes are due to the well-known pump-induced inward-shifting of polariton modes known as Rabi contraction [6], [35]. Rabi contraction results simply from a reduction in the collective coupling strength as molecular population is transiently driven out of the lower state of the cavity-coupled transition. The timescales for the appearance and decay of these Rabi contraction features are therefore similar to those of the extracavity GSB dynamics.

The ESA dynamics are slightly more difficult to see in the transient intracavity data. We know from the extracavity results (Section 3.2) that the ESA of the carbonyl mode of interest blueshifts from ~2040 cm$^{-1}$ past 2060 cm$^{-1}$ in the first 5–10 ps following pump excitation. This ESA feature therefore passes through the spectral region of the UP formed from strong cavity coupling of the ground state carbonyl vibration, leading to a transient reduction in cavity transmission in the UP region. The ESA therefore manifests as increased ΔOD in the UP region (near 2040 cm$^{-1}$) as compared to the LP region (near 2000 cm$^{-1}$) in the early time traces in Fig. 4H. This overlap of the ESA with the ground-state UP is precisely what will allow us to reconstruct the intracavity ESA dynamics (see Section 3.4 below). At the same time, the ESA/UP overlap may ultimately present a challenge for achieving excited-state VSC in this system, which may well require a larger shift between ground and excited state vibrational frequencies to avoid spectral congestion.

In any event, this qualitative discussion of transient cavity transmission signals is insufficient for accurate extraction of the dynamics of the intracavity molecules. Transient cavity spectra are more complex than transient absorption spectra because they bake in the response of the changing optical interference conditions along with the dynamics of the intracavity molecules. It is therefore not sound practice to take temporal lineouts of differential cavity transmission traces to analyze population dynamics as one would for a GSB or ESA feature; the positive and negative lobes of, e.g., the derivative lineshapes in Fig. 4H simply do not correlate directly with state populations. Changes in cavity transmission at a given frequency arise from both shifting positions of cavity modes, as well as changes in intracavity absorption at that frequency. For instance, a temporal lineout of the transient cavity transmission data at the 2040 cm$^{-1}$ UP frequency will feature overlapping dynamics from both the Rabi contraction and absorption from the shifting ESA feature. One must be thoughtful about disentangling these effects. We therefore now turn to fitting the transient cavity transmission spectra to reconstruct the intracavity molecular response, an observable that can be compared directly to the extracavity control measurements.

### 3.4 Spectral reconstruction of the intracavity molecular response

We now introduce a spectral reconstruction algorithm that we use to extract the intrinsic intracavity molecular response from transient cavity transmission spectra. This procedure is depicted in Fig. 5 and consists of two steps performed for the transient cavity transmission spectrum acquired at each time delay: (a) fitting the pump-off cavity transmission spectrum and (b) fitting the differential cavity transmission spectrum. In each of these steps, we rely on the analytical expression for the frequency-dependent fractional transmission of light, $I_T(v)/I_0$, through a Fabry-Pérot cavity composed of two identical mirrors [8], [37], [40], [41]:

$$\frac{I_T(v)}{I_0} = \frac{T^2 e^{-\alpha(v)L}}{1+R^2 e^{-2\alpha(v)L}-2Re^{-\alpha(v)L}\cos\left[\frac{4\pi Ln(v)v}{c}\right]} \quad (1)$$

Here $v$ is frequency, $T$ and $R$ are the transmission and reflection intensity coefficients for each mirror, $\alpha(v)$ and $n(v)$ are the frequency-dependent absorption coefficient and refractive index of the intracavity medium, $L$ is the cavity length, and $c$ is the speed of light.

In the first step of the routine (Fig. 5A), we construct the pump-off (unperturbed) transmission spectrum for the cavity-coupled ReCl(CO)₃(bpy) system. Here, we work only in the narrow spectral region around the $a'$ symmetric carbonyl stretch. We define the unperturbed frequency-dependent absorption coefficient of intracavity molecules, $\alpha_0(v)$, as a single Lorentzian peak with amplitude $c_0$, center frequency $v_0$, and fwhm linewidth $\gamma_0$. An initial guess for $\alpha_0(v)$ is generated by fitting the extracavity static absorption spectrum to constrain $c_0$, $v_0$, and $\gamma_0$. From this initial guess for $\alpha_0(v)$, we compute the corresponding initial refractive index $n_0(v)$ via the Kramers-Kronig relation. Initial guesses for $T$ and $R$ are taken from experimental measurements of the ITO mirrors, and the cavity length $L$ is initially taken to be 25 μm. We plug these initial values for $\alpha_0(v)$, $n_0(v)$, $T$, $R$, and $L$ into Eq. 1 to generate an initial guess for the pump-off transmission spectrum $I_{T,0}(v)$. We then use a non-linear least-squares optimization (fmincon method in MATLAB) to refine the values of $c_0$, $v_0$, $\gamma_0$, $T$, $R$, and $L$ by minimizing the residual between the simulated $I_{T,0}(v)$ and the experimental pump-off spectrum. This procedure is repeated for the pump-off spectrum acquired at each time delay to define the reference point for subsequent fitting of the transient differential spectrum at each time delay.

In the second step of the routine (Fig. 5B), we construct the differential transmission spectrum for the cavity-coupled system at time $t$ following optical pumping. We define the time-dependent intracavity absorption coefficient $\alpha_t(v) = \alpha_0(v) + \Delta\alpha_t(v)$, where the initial guess for $\alpha_0(v)$ is obtained from step 1, and $\Delta\alpha_t(v)$ captures the transient absorption of intracavity molecules at time $t$. For ReCl(CO)₃(bpy), we construct $\Delta\alpha_t(v)$ as a sum of three Lorentzian components to capture the central GSB of the $a'$ symmetric carbonyl stretch and two ESA features, one on either side of the GSB. We find that adding additional Lorentzian components does not significantly improve the fitting results or alter the resulting extracted dynamics in this system.

We generate initial guesses for the amplitudes $\{c_n\}$, center frequencies $\{v_n\}$, and fwhm linewidths $\{\gamma_n\}$ ($n$ = 1–3) of the Lorentzian components of $\Delta\alpha_t(v)$ by fitting the extracavity transient absorption data at the same time delay. We compute $\alpha_t(v)$ from the initial guesses for $\alpha_0(v)$ and $\Delta\alpha_t(v)$, then use the Kramers-Kronig relation to obtain the corresponding time-dependent refractive index $n_t(v)$. We simulate an initial guess for $I_{T,t}(v)$, the pump-on cavity transmission spectrum at time $t$, by plugging in our guesses for $\alpha_t(v)$ and $n_t(v)$ into Eq. 1, along with the values of $T$, $R$, and $L$ obtained in step 1. Finally, an initial guess for the differential cavity transmission spectrum is calculated according to $\Delta I_{T,t}(v) = -\log_{10}[\ I_{T,t}(v) / I_{T,0}(v)\ ]$, using the initial guess for $I_{T,t}(v)$ and the pump-off transmission spectrum $I_{T,0}(v)$ from step 1. An iterative optimization loop (again implemented with the MATLAB fmincon method) subsequently refines the values of $c_0$, $v_0$, $\gamma_0$, $\{c_n\}$, $\{v_n\}$, and $\{\gamma_n\}$ until the simulated $\Delta I_{T,t}(v)$ converges with the experimental differential cavity transmission spectrum. Note that we allow the values of $c_0$, $v_0$, and $\gamma_0$ determined in step 1 to float in step 2, which impact both $I_{T,t}(v)$ and $I_{T,0}(v)$ in the calculation of $\Delta I_{T,t}(v)$. The fitted values for $c_0$, $v_0$, and $\gamma_0$ found in step 2 ultimately do not deviate substantially from those found in step 1. We have also experimented with refining the mirror transmission and reflection coefficients ($T$, $R$) and cavity length $L$ again in step 2, but find that floating these parameters does not noticeably improve the fits. We therefore leave the $T$, $R$, and $L$ parameters fixed at the values determined in step 1.

The central outcome of this fitting routine is to determine the transient intracavity absorption coefficient, $\Delta\alpha_t(v)$. Importantly, $\Delta\alpha_t(v)$ disentangles the time-dependent molecular response from the obscuring optical filtering of the cavity and can be directly compared to extracavity transient absorption data. To demonstrate this, we plot representative reconstructed intracavity transient absorption spectra for ReCl(CO)₃(bpy) in the right-hand column of Fig. 4, obtained from fitting the intracavity data in the central column of Fig. 4. The $\alpha_0(v)$ absorption coefficient used to fit the pump-off spectrum is shown in Fig. 4C. The reconstructed $\Delta\alpha_t(v)$ spectra are plotted in Fig. 4F as a function of wavenumber and pump-probe delay, and lineouts at selected time delays are plotted in Fig. 4I.

The reconstructed intracavity molecular response under VSC (Figs. 4FI) closely resembles the behavior observed in the extracavity data (Figs. 4DG), recovering both the GSB and ESA features. To quantify the dynamics of these features, we take temporal lineouts from the reconstructed $\Delta\alpha_t(v)$ spectra and perform exponential fits to extract rise and decay time constants just as we did for the extracavity data. Details are provided in Section S1 of the Supplementary Material. Extracted time constants from all experiments performed under resonant VSC conditions are laid out in Table S2 and values averaged across all experiments are summarized in Table 1. The spectral reconstruction approach also robustly extracts intracavity molecular dynamics in cavities detuned from resonance, as presented in Section S2 of the Supplementary Material. Extracted time constants from all experiments performed in detuned cavities are laid out in Table S3, and average values are again summarized in Table 1. All time constants measured under both resonant VSC and in detuned cavities are found to be statistically indistinguishable from the extracavity results, confirming that ground-state vibrational cavity-coupling does not detectably alter the excited state dynamics of ReCl(CO)$_3$(bpy).

We close this section by noting that the spectral reconstruction approach introduced here is complementary to other methods used in the polariton spectroscopy literature. A more common strategy is "forward construction" of transient cavity transmission data by considering optical cavity filtering of extracavity transient absorption data; the work of Lüttgens *et al*. [42] is a good example. In Section S3 of the Supplementary Material, we demonstrate that we can indeed accurately simulate transient cavity transmission data by processing extracavity transient absorption data using Eq. 1. This provides further validation that extracavity and intracavity ReCl(CO)$_3$(bpy) molecules feature indistinguishable dynamics. That said, our reconstruction algorithm provides a cleaner, more interpretable means to compare extracavity and intracavity dynamics, as the evolution of features in transient absorption spectra are easier to interpret and quantify than transient cavity transmission spectra, per our discussion in Section 3.3.

## 4 Conclusions

The central outcome of this work is the optical modulation of vibrational strong coupling in intracavity ReCl(CO)$_3$(bpy) molecules via excitation with UV light. The carbonyl stretching region of ReCl(CO)$_3$(bpy) proves an apt system for this aim, as the excited state vibrational modes are blue-shifted from the ground state modes, yielding well-separated GSB and ESA features in the transient spectra.

We introduced a spectral reconstruction protocol to extract the intrinsic dynamics of intracavity molecules from transient cavity transmission spectra. We expect that this will be an important tool for the spectroscopy of polaritonic systems moving forward. By parameterizing both static and transient intracavity absorption coefficients in terms of Lorentzian components, then propagating these quantities through the FP expression for cavity transmission, we obtain a clear picture of the intracavity molecular response that circumvents cavity-mediated optical filtering effects. The time constants we extract for the excited-state vibrational dynamics of ReCl(CO)$_3$(bpy) are consistent across all intracavity and extracavity experiments within experimental error. This finding underscores the quantitative performance of our reconstruction method and confirms that ground-state VSC does not alter excited state dynamics in this system. While we only showcase reconstruction of the transient UV-pump/IR-probe spectra of ReCl(CO)$_3$(bpy) here, we believe that this approach can be easily generalized for quantitative studies of the dynamics of a range of molecular systems under cavity strong coupling.

Our spectral reconstruction protocol highlights the fact that cavity coupling can yield a significant enhancement of transient signals. In the UV-pump/IR-probe measurements of ReCl(CO)$_3$(bpy) embedded in ITO FP cavities, we observe differential cavity transmission signals as large as 15–20 ΔmOD that arise from transient changes in the reconstructed intracavity molecular absorption coefficient of only ∼5 ΔmOD. We therefore quote a roughly 3–4-fold cavity-enhancement of transient signals in the cavities used here. This signal increase is consistent across the measurements performed both under resonant VSC (comparing the peak transient signal excursion in Fig. 4EH to those in Fig. 4FI) as well as under detuned intracavity conditions (comparing the peak transient signal excursion in Fig. S5EH to those in Fig. S5FI). We ascribe this signal enhancement to classical optical cavity-enhanced effects, which we have discussed in more detail elsewhere [37]. Put simply, optical cavities are extremely sensitive to the complex refractive index of the intracavity medium, and their transmission spectra can accordingly amplify tiny changes in intracavity absorption. We anticipate that this cavity-mediated amplification of transient signals may be broadly useful for improving detection limits in condensed-phase spectroscopy, as has already been explored in the context of ultrafast gas-phase dynamics [43].

This effort also represents a first step towards potential implementations of excited-state VSC and polariton-mediated IR-UV photonic transduction. Per the recent proposal of Waverly *et al*. [26], IR-UV transduction requires cavity-coupling of a vibrational mode that is both IR and Franck-Condon active; the $a'$ carbonyl stretch of ReCl(CO)$_3$(bpy) targeted for VSC here is of the right symmetry to meet these conditions. However, based on our findings here, the excited state vibrational absorption bands of ReCl(CO)$_3$(bpy) may ultimately prove too weak – and perhaps too close in frequency to the corresponding ground-state bands – for excited-state VSC. It is worth continuing the search for molecular candidates that feature stronger IR transition dipoles and more brightly absorbing optical transitions, or that can be prepared in denser samples. If such systems can be identified, it will be of great interest to extend the ultrafast platform we introduce here to examine excited-state dynamics under VSC and test theoretical predictions for how cavity-coupling of excited-state vibrations might impact photophysics.

## Authors' statements

**Supplementary Material:** See the Supplementary Material for temporal fits of ground-state bleach and excited-state absorption dynamics from extracavity and reconstructed intracavity datasets; control measurements under detuned conditions; representative forward-construction of transient cavity transmission spectra from extracavity data; and reported time constants from independent experimental datasets.

**Acknowledgements:** We thank Professor Haochuan Mao and Professor Wei Xiong for helpful discussion regarding ITO mirror design.

**Research funding:** This work was supported by the US Department of Energy, Office of Science, Basic Energy Sciences, CPIMS Program under Early Career Research Program award DE-SC0022948.

**Author contributions:** All authors have accepted responsibility for the entire content of this manuscript and consented to its submission to the journal, reviewed all the results and approved the final version of the manuscript.

**Conflict of interest:** The authors state no conflicts of interest.

**Data availability statement:** The datasets generated and/or analyzed during the current study are available from the corresponding author upon reasonable request.

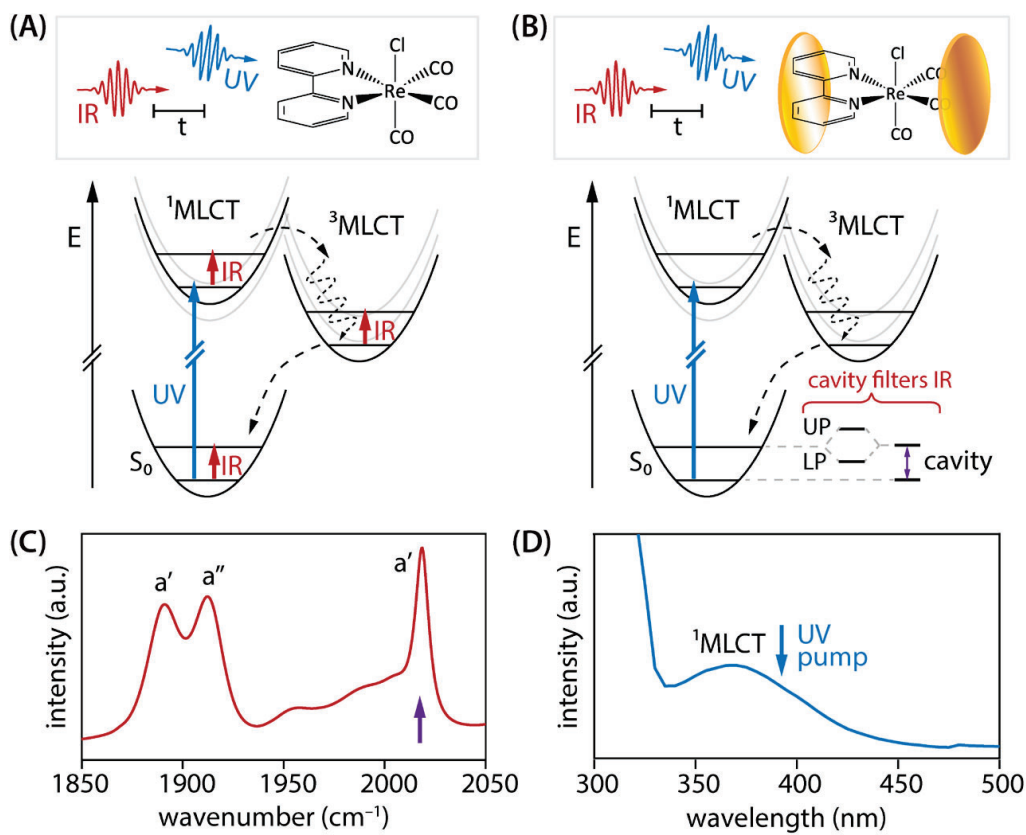

**Figure 1** | **(A)** Experimental scheme for ultrafast UV-pump/IR-probe spectroscopy of ReCl(CO)$_3$(bpy). Excitation with a 390 nm UV pump pulse initiates excited-state dynamics, which are subsequently probed with a broadband mid-IR probe pulse centered near 2000 cm$^{-1}$. The schematic energy-level diagram illustrates how photoexcitation drives ReCl(CO)$_3$(bpy) from its S$_0$ electronic ground state to the singlet metal-to-ligand charge transfer ($^1$MLCT) manifold, followed by intersystem crossing to the triplet ($^3$MLCT) manifold, and ultimately vibrational cooling and relaxation back to the ground state. **(B)** Experimental scheme for spectroscopy of ReCl(CO)$_3$(bpy) under vibrational strong coupling (VSC) in a microfluidic Fabry-Pérot cavity. Resonant VSC of a carbonyl stretching mode in the S$_0$ ground state generates upper and lower polariton states (UP, LP). Excitation with a UV pump pulse again initiates excited-state dynamics, while a broadband mid-IR probe pulse records subsequent transient changes to the cavity transmission spectrum in the carbonyl stretching region. **(C)** Infrared absorption spectrum of ReCl(CO)$_3$(bpy) in solution in DMSO in the carbonyl stretching region. The $a'$ symmetric carbonyl stretch near 2018 cm$^{-1}$ (marked with a purple arrow) is targeted for resonant VSC in this work. **(D)** UV-visible absorption spectrum of ReCl(CO)$_3$(bpy) in solution in DMSO. The 390 nm pump used here excites the red shoulder of the broad $^1$MLCT band.

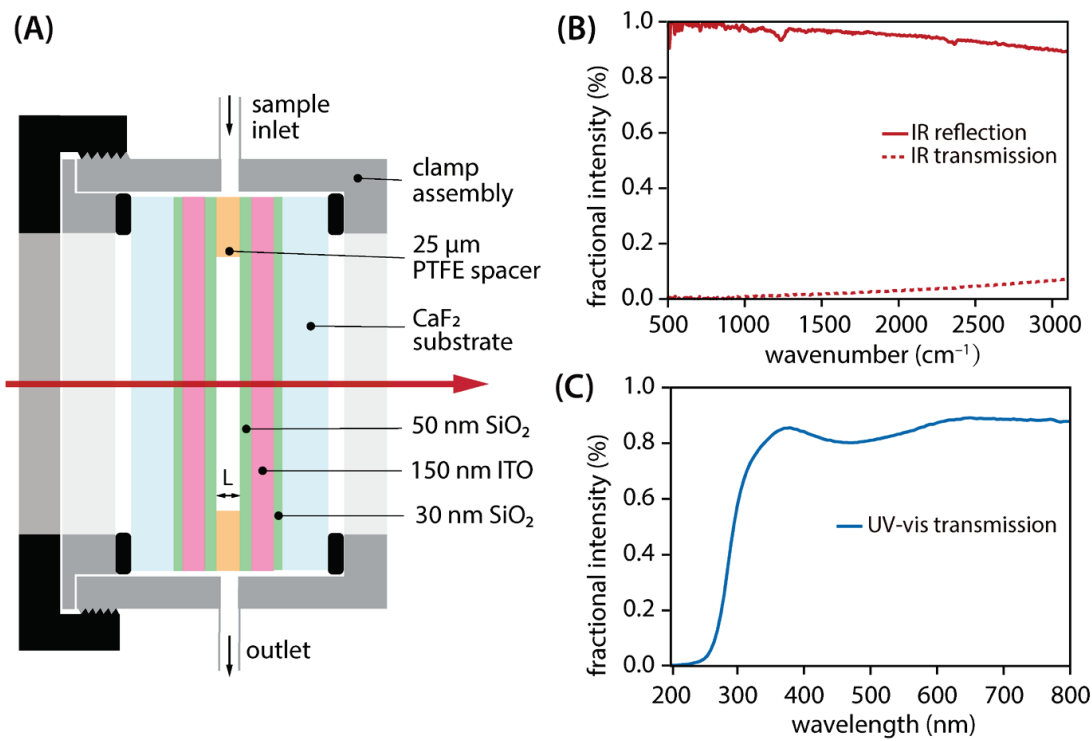

**Figure 2 |** **(A)** Schematic of microfluidic Fabry-Pérot cavity used here for vibrational strong coupling and UV-pump/IR-probe experiments. The cavity is assembled from two indium tin oxide (ITO)-coated CaF$_2$ substrates separated by a 25 μm polytetrafluoroethylene (PTFE) spacer. **(B)** Reflectivity and transmission spectra of a single ITO mirror confirming high reflectivity across the mid-IR. **(C)** Transmission spectrum of a single ITO mirror demonstrating high broadband transmittance in the UV-visible.

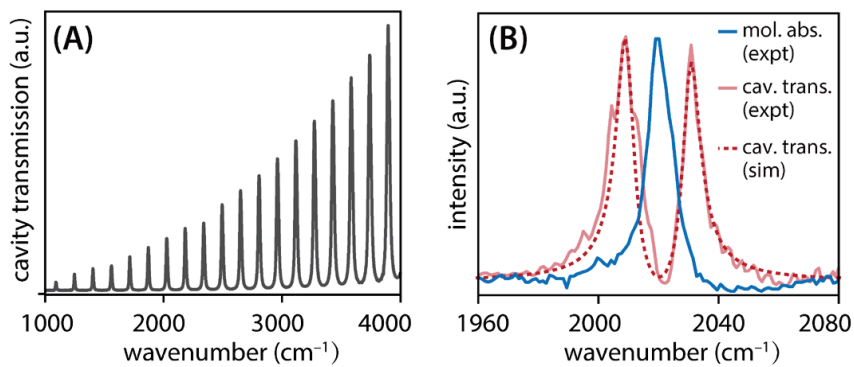

**Figure 3 |** **(A)** Transmission spectrum of a representative empty Fabry-Pérot cavity formed by two ITO mirrors separated by a 25 μm PTFE spacer. **(B)** Cavity transmission spectrum of ReCl(CO)$_3$(bpy) in DMSO under strongly-coupled conditions in a 25 μm microfluidic cavity (solid light red) plotted against the bare molecular absorption spectrum in the region near the *a′* symmetric carbonyl stretch of ReCl(CO)$_3$(bpy) (blue). A simulated strongly-coupled transmission spectrum is also provided (red dotted lines) using the classical expression for the transmission spectrum of a Fabry-Pérot cavity (Eq. 1).

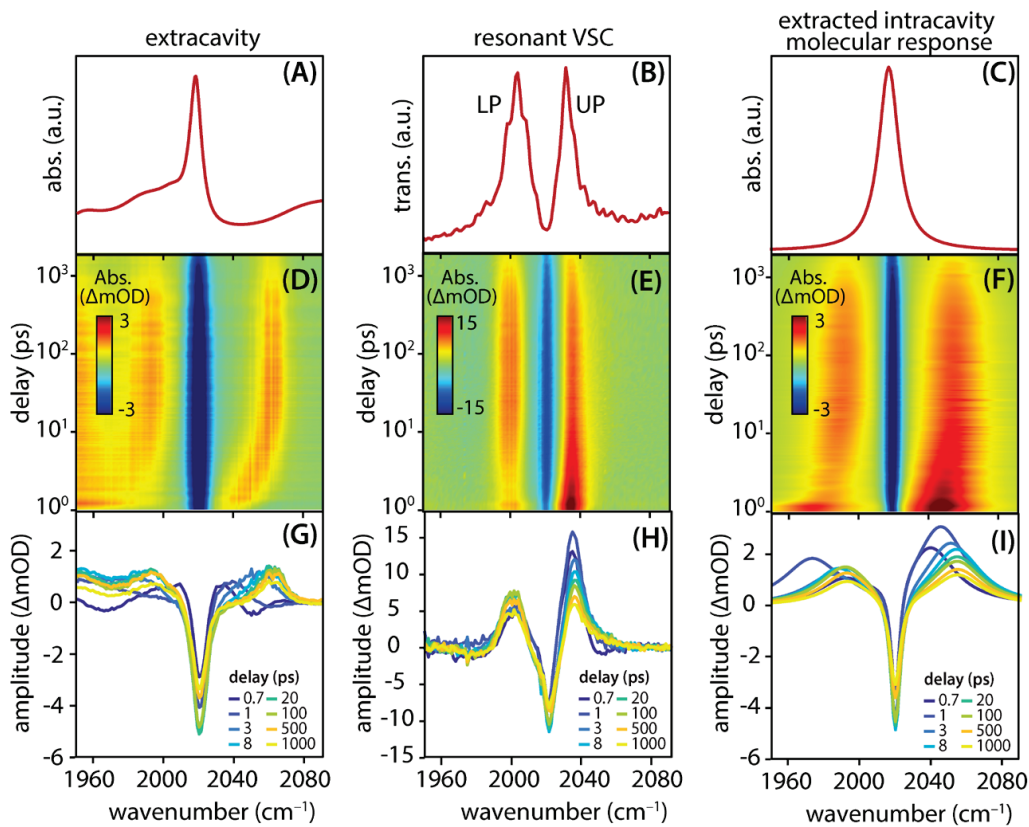

**Figure 4** | Ultrafast UV-pump/IR-probe spectroscopy of ReCl(CO)$_3$(bpy) in DMSO. Extracavity control data is presented in the left-hand column and intracavity data collected under resonant VSC conditions is presented in the central column. The right-hand column presents the molecular response extracted from intracavity data, which can be compared directly to extracavity data. **(A)** Steady-state extracavity IR absorption spectrum of ReCl(CO)$_3$(bpy) in DMSO showing the $a'$ symmetric carbonyl stretching mode near 2018 cm$^{-1}$. **(B)** Static (pump-off) transmission spectrum of strongly-coupled cavity filled with ReCl(CO)$_3$(bpy) in DMSO, exhibiting upper and lower polariton (UP, LP) features with a Rabi splitting of ∼24 cm$^{-1}$. **(C)** Representative pump-off intracavity absorption spectrum for ReCl(CO)$_3$(bpy), $\alpha_0(\nu)$, modeled with a single Lorentzian lineshape and fit to the pump-off cavity transmission spectrum in the first step of the spectral reconstruction algorithm. **(D,E,F)** Ultrafast UV-pump/IR-probe spectra plotted as a function of delay time and wavenumber showing **(D)** the differential absorption of an extracavity sample, **(E)** the differential cavity transmission of an intracavity sample under resonant VSC, and **(F)** the intracavity molecular response, $\Delta\alpha_t(\nu)$, reconstructed from the data in panel (E). **(G, H, I)** Spectral lineouts at selected pump-probe delays of transient data acquired for **(G)** the extracavity sample, **(H)** the intracavity sample, and **(I)** the reconstructed intracavity molecular response.

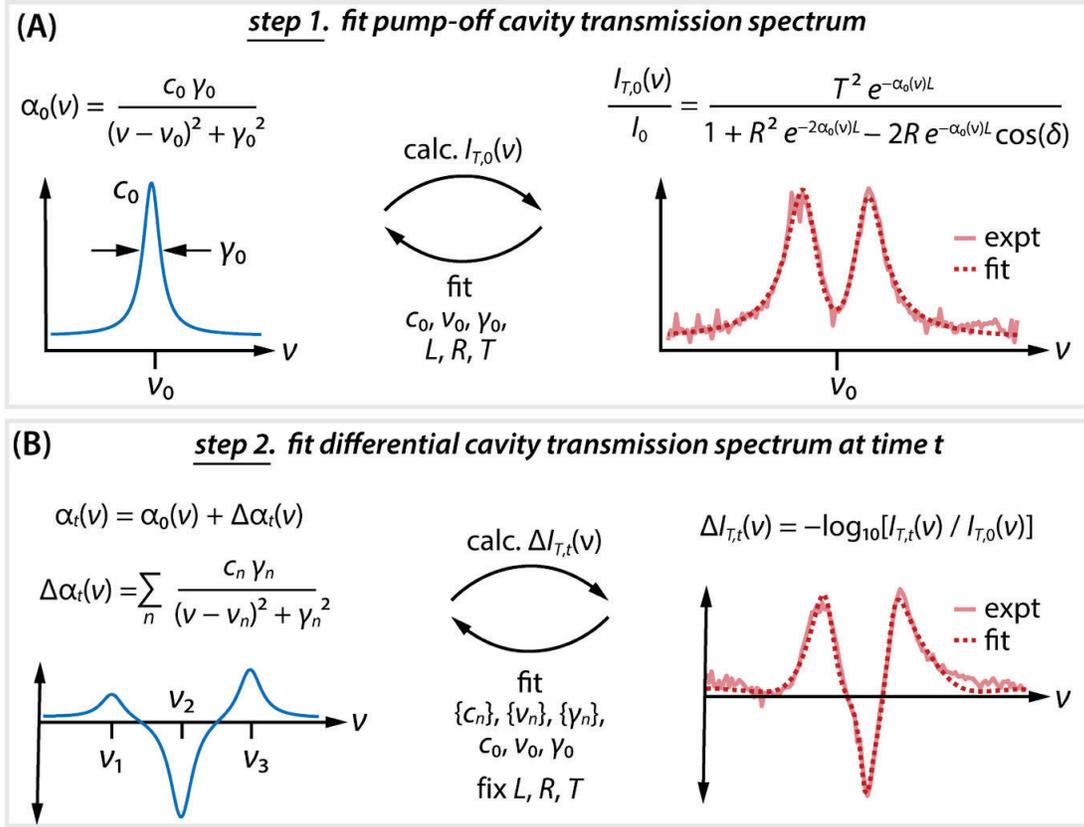

**Figure 5** | Spectral reconstruction workflow used to extract intracavity molecular dynamics from strongly-coupled cavity transmission data. **(A)** In step 1, we fit the pump-off transmission spectrum for the cavity-coupled system, $I_{T,0}(\nu)$. We represent the pump-off intracavity molecular absorption coefficient, $\alpha_0(\nu)$, as a single Lorentzian function with amplitude $c_0$, central frequency $\nu_0$, and fwhm linewidth $\gamma_0$. We perform a nonlinear fit between simulated and experimental pump-off spectra by floating $c_0$, $\nu_0$, and $\gamma_0$, along with the cavity length $L$, mirror reflectivity $R$, and mirror transmission $T$. **(B)** In step 2, we fit the differential cavity transmission spectrum acquired at time $t$, $\Delta I_{T,t}(\nu)$. The time-dependent intracavity absorption coefficient, $\alpha_t(\nu) = \alpha_0(\nu) + \Delta\alpha_t(\nu)$, is constructed by taking the transient absorption $\Delta\alpha_t(\nu)$ as a sum of three Lorentzian functions with amplitudes $\{c_n\}$, center frequencies $\{\nu_n\}$, and fwhm widths $\{\gamma_n\}$. We use $\alpha_t(\nu)$, along with the parameters $L$, $R$, and $T$ optimized in step 1, to generate the pump-on cavity transmission spectrum for a given pump-probe time delay, $I_{T,t}(\nu)$. The differential cavity transmission spectrum $\Delta I_{T,t}(\nu)$ is then constructed from $I_{T,t}(\nu)$ and $I_{T,0}(\nu)$. A second nonlinear optimization loop refines both the transient Lorentzian parameters ($\{c_n\}$, $\{\nu_n\}$, and $\{\gamma_n\}$) and the pump-off Lorentzian parameters ($c_0$, $\nu_0$, and $\gamma_0$) to best reproduce the experimental differential cavity transmission spectrum. This process enables extraction of $\Delta\alpha_t(\nu)$, which encodes the intrinsic dynamics of the intracavity molecules.

**Table 1** | Time constants for the ground-state bleach (GSB) rise and decay and excited-state absorption (ESA) decay and shift from UV-pump/IR-probe experiments on ReCl(CO)$_3$(bpy) in DMSO. Values are reported for extracavity control samples as well as intracavity samples under resonant VSC of the *a′* symmetric carbonyl stretch at 2018 cm$^{-1}$ and detuned intracavity control samples. Error bars represent standard deviations arising from averaging time constants from the individual datasets presented in Tables S1–S3 in Section S4 of the Supplementary Material.

| Dynamic process | Extracavity | Intracavity, resonant VSC | Intracavity, detuned |
|---|---|---|---|
| GSB rise (ps) from linecut at 2018 cm$^{-1}$ | 1.4 ± 0.3 | 1.6 ± 0.4 | 1.6 ± 0.3 |
| GSB decay (ps) from linecut at 2018 cm$^{-1}$ | 150 ± 10 | 170 ± 20 | 150 ± 40 |
| ESA decay (ps) from linecut at 2060 cm$^{-1}$ | 120 ± 20 | 150 ± 30 | 130 ± 30 |
| ESA shift (ps) from Lorentzian fit | 6.1 ± 0.7 | 5.2 ± 1.2 | 6.0 ± 1.6 |

# Ultrafast optical modulation of vibrational strong coupling in ReCl(CO)$_3$(2,2-bipyridine)

Liying Chen,[1] Alexander M. McKillop,[1] Ashley P. Fidler,[1,2] and Marissa L. Weichman[1,*]

[1]Department of Chemistry, Princeton University, Princeton, New Jersey 08544, USA

[2]Present Address: Chemistry Division, Naval Research Laboratory, Washington, DC 20375, USA

[*]weichman@princeton.edu

**Section S1: Temporal fitting of extracavity and intracavity vibrational dynamics**

We perform time-domain fitting of the extracavity transient data for ReCl(CO)$_3$(2,2-bipyridine) discussed in Section 3.2 of the main text, as well as the intracavity $\Delta\alpha_t(v)$ datasets extracted using the spectral reconstruction algorithm detailed in Section 3.4. Temporal lineouts are taken at characteristic wavenumbers corresponding to the ground-state bleach (GSB, 2018 cm$^{-1}$) and the excited-state absorption (ESA, 2060 cm$^{-1}$). For the shifting ESA band, we additionally fit the feature with a Lorentzian function at each delay time in order to extract the center frequency as a function of time. The temporal evolution of the lineouts and frequency shifts are fit using exponentials to determine rise, decay, and shift time constants. Representative exponential fits are shown in Fig. S1 for the GSB rise dynamics, Fig. S2 for the GSB decay dynamics, Fig. S3 for the ESA decay dynamics, and Fig. S4 for the ESA shift. The left-hand panels of Figs. S1–S4 show fitting of the same representative extracavity dataset shown in the left-hand column of Fig. 4 in the main text. The right-hand panels of Figs. S1–S4 show fits for the same reconstructed intracavity dataset under resonant VSC shown in the right-hand column of Fig. 4. Extracted time constants averaged across all data sets are summarized in Table 1 of the main text.

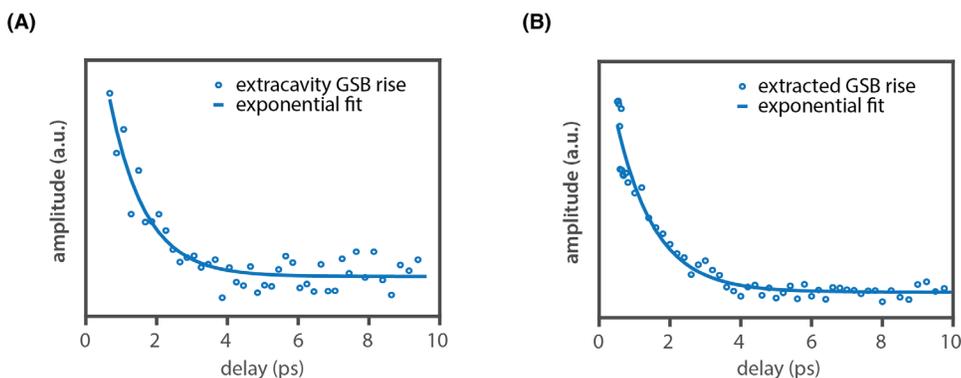

**Figure S1** | Early time ground-state bleach (GSB) rise dynamics from lineouts taken at 2018 cm$^{-1}$ in representative **(A)** extracavity data from the left-hand column of Fig. 4 and **(B)** intracavity data collected under resonant VSC and reconstructed using the algorithm detailed in Section 3.4 of the main text. Experimental data is plotted in points, while solid lines represent best-fit curves. To fit the early-time GSB rise, data points for $t < 10$ ps are fit with a single exponential function.

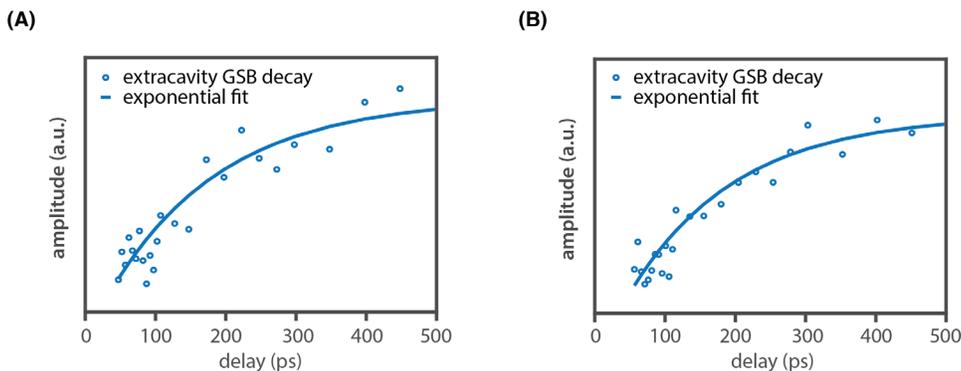

**Figure S2** | Long-time ground-state bleach (GSB) decay dynamics from lineouts at 2018 cm$^{-1}$ in representative **(A)** extracavity data from the left-hand column of Fig. 4 and **(B)** intracavity data collected under resonant VSC and reconstructed using the algorithm detailed in Section 3.4 of the main text. Experimental data is plotted in points, while solid lines represent best-fit curves. To fit the long-time GSB decay, data points for $t > 50$ ps are fit with a single exponential function.

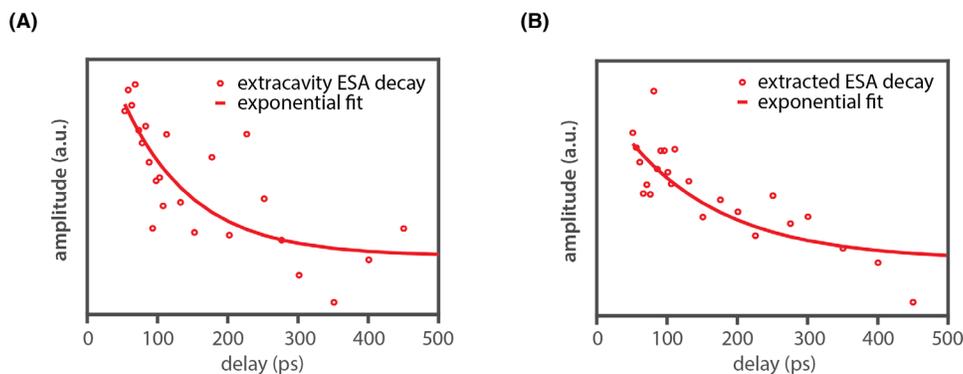

**Figure S3** | Long-time excited state absorption (ESA) decay dynamics from lineouts at 2060 cm$^{-1}$ in representative **(A)** extracavity data from the left-hand column of Fig. 4 and **(B)** intracavity data collected under resonant VSC and reconstructed using the algorithm detailed in Section 3.4 of the main text. Experimental data is plotted in points, while solid lines represent best-fit curves. To fit the long-time ESA decay, data points for $t > 50$ ps are fit with a single exponential function.

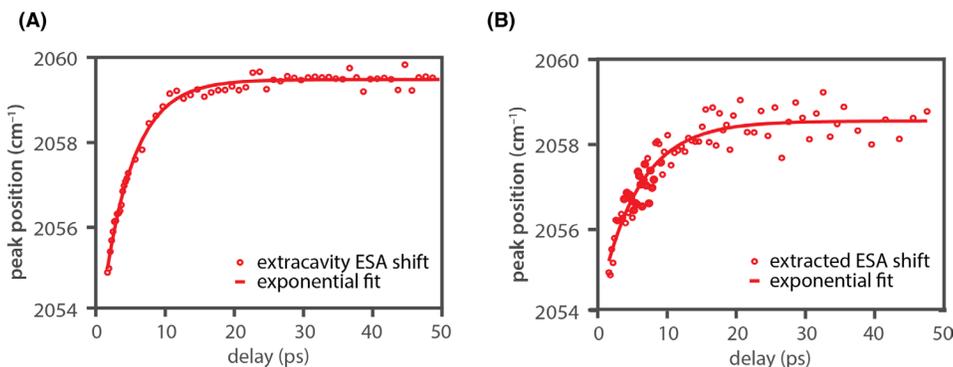

**Figure S4** | Excited-state absorption (ESA) peak shift dynamics obtained from Lorentzian fitting of the ESA peak center as a function of pump-probe delay time in **(A)** extracavity data from the left-hand column of Fig. 4 and **(B)** intracavity data collected under resonant VSC and reconstructed using the algorithm detailed in Section 3.4 of the main text. Experimental data is plotted in points, while solid lines represent best-fit curves. Data points for $t < 50$ ps are fit with a single exponential function.

**Section S2: Detuned cavity control experiments**

To examine role of cavity detuning, we perform control experiments on intracavity ReCl(CO)$_3$(bpy) in microfluidic FP cavities where the cavity length is adjusted away from exact resonance with the *a′* symmetric carbonyl stretch at 2018 cm$^{-1}$. We access these detuning conditions by laterally translating the cavity in the plane of the mirrors, taking advantage of slight deviations from perfect mirror parallelism to vary the effective cavity length at the location where the pump-probe measurement is performed.

Data from a representative UV-pump/IR-probe experiment performed for intracavity ReCl(CO)$_3$(bpy) under detuned conditions are shown in Fig. S5. The left-hand column of Fig. S5 reproduces the same extracavity transient data as the left-hand column of Fig. 4 of the main text. The central column of Fig. S5 shows the raw transient cavity transmission data for the detuned device. Under these conditions, the static pump-off cavity transmission spectrum (Fig. S5B) exhibits two polaritonic peaks with unequal amplitudes and a splitting different from the resonant VSC case. The raw transient transmission spectra for the detuned cavity (Fig. S5EH) display derivative-like lineshapes rather than simple bleach or absorption features. The intracavity molecular response extracted with our spectral reconstruction algorithm is plotted in the right-hand column of Fig. S5. The reconstructed spectra clearly resolve the underlying molecular response, yielding clear GSB and ESA features. The extracted time constants for the temporal evolution of these features under detuned cavity-coupling conditions are summarized in Table 1 of the main text and are consistent with the extracavity control.

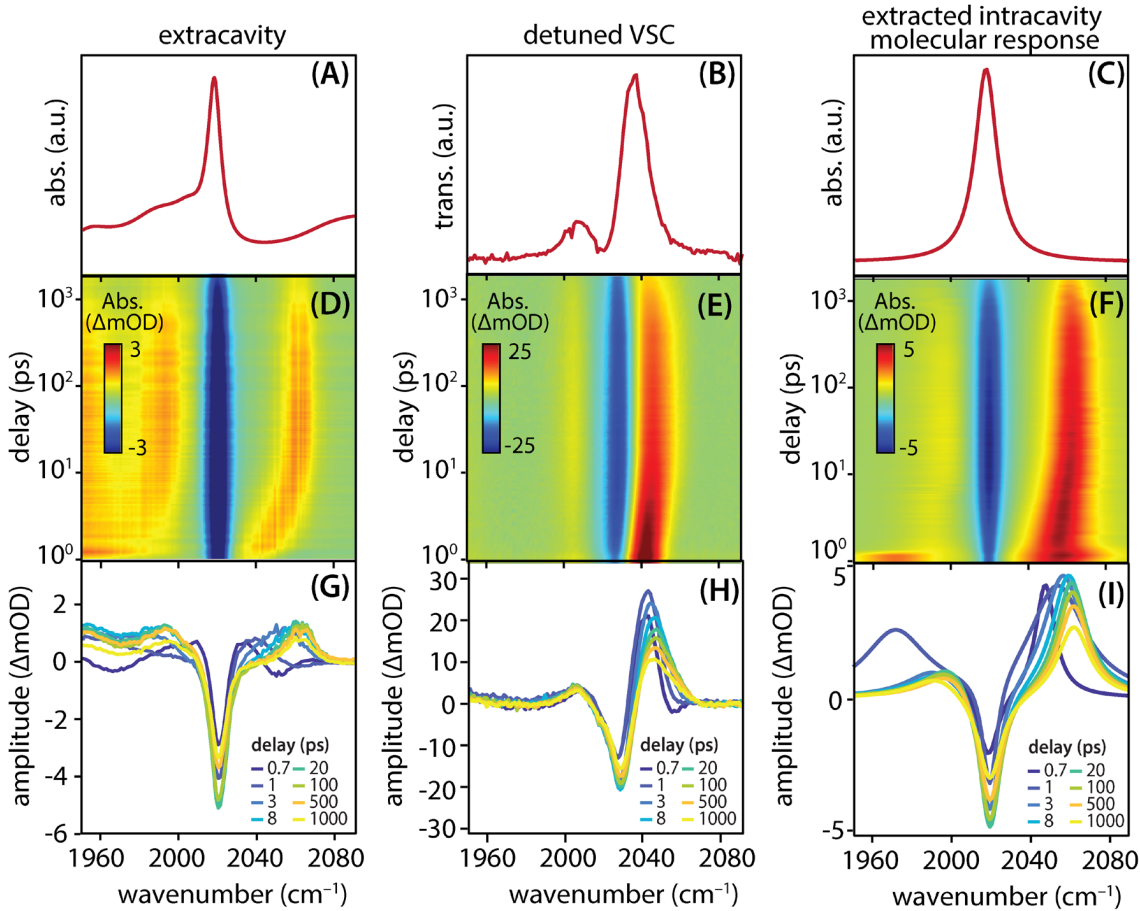

**Figure S5** | Ultrafast UV-pump/IR-probe spectroscopy of ReCl(CO)$_3$(bpy) in DMSO. Extracavity control data is presented in the left-hand column (reproduced from Fig. 4) and intracavity data collected under detuned VSC conditions is presented in the central column. The right-hand column presents the molecular response extracted from intracavity data, which can be compared directly to extracavity data. **(A)** Steady-state extracavity IR absorption spectrum of ReCl(CO)$_3$(bpy) in DMSO showing the $a'$ symmetric carbonyl stretching mode near 2018 cm$^{-1}$. **(B)** Static (pump-off) transmission spectrum of a detuned cavity filled with ReCl(CO)$_3$(bpy) in DMSO, exhibiting asymmetric polariton features. **(C)** Representative pump-off intracavity absorption spectrum for ReCl(CO)$_3$(bpy), $\alpha_0(\nu)$, modeled with a single Lorentzian line-shape and fit to the pump-off cavity transmission spectrum in the first step of the spectral reconstruction algorithm. **(D,E,F)** Ultrafast UV-pump/IR-probe spectra plotted as a function of delay time and wavenumber showing **(D)** the differential absorption of an extracavity sample, **(E)** the differential cavity transmission of an intracavity sample under detuned coupling conditions, and **(F)** the intracavity molecular response, $\Delta\alpha_t(\nu)$, reconstructed from the data in panel (E). **(G, H, I)** Spectral lineouts at selected pump-probe delays of transient data acquired for **(G)** the extracavity sample, **(H)** the intracavity sample, and **(I)** the reconstructed intracavity molecular response.

## Section S3: Forward-construction of transient cavity transmission spectra

In the main text, we emphasize reconstruction of the intracavity transient absorption from the transient cavity transmission spectra, which can then be directly compared to extracavity control data. One can also "forward-construct" the transient cavity transmission spectrum by applying the Fabry-Pérot expression in Eq. (1) of the main text to extracavity transient absorption data. The constructed cavity transmission spectra can then be compared against raw cavity data. Figure S6 provides an example of how we also find good agreement between extracavity and intracavity spectra using this approach. Note, however, that interpretation of transient cavity spectra are more nuanced than transient absorption spectra, as discussed in Section 3.3 of the main text, making it more involved to interpret the features of these data than it is to understand the reconstructed intracavity molecular response data (Fig. 4).

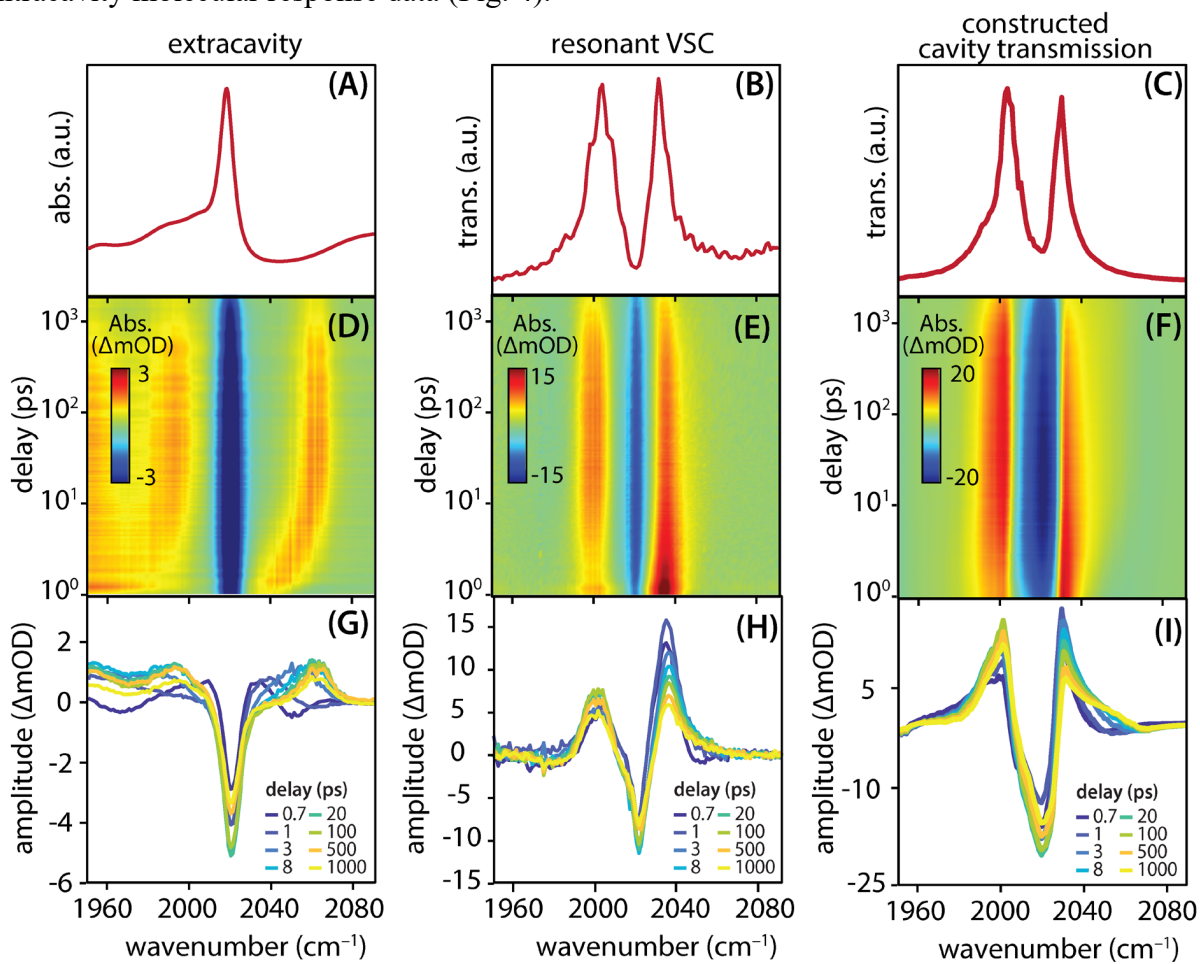

**Figure S6** | Ultrafast UV-pump/IR-probe spectroscopy of ReCl(CO)$_3$(bpy) in DMSO. **(A, D, G)** Extracavity data and **(B, E, H)** resonant intracavity data reproduced from Fig. 4. **(C, F, I)** Transient cavity transmission spectra constructed by processing the extracavity data in the left-hand column with Eq. (1) of the main text. **(C)** Simulated pump-off transmission spectrum of a 25.07 μm-long cavity with $R = 0.95$ mirrors, filled with dielectric material featuring an absorption coefficient given by the experimental extracavity data in panel (A). **(F)** Simulated differential UV-pump/IR-probe cavity transmission spectra plotted as a function of delay time and wavenumber. Transient traces are constructed by calculating pump-on and pump-off cavity transmission spectra from the corresponding pump-on and pump-off extracavity data that underlie the spectra in panel (D). **(I)** Spectral lineouts at selected pump-probe delays of transient data from panel (F).

S6

## Section S4: Reproducibility of transient extracavity and intracavity measurements

We collect multiple independent transient datasets for each condition (extracavity, intracavity under resonant VSC, and intracavity with detuned coupling) and extract time constants from each dataset using the fitting methods described in the main text and in Section S1. The results from independent datasets are summarized in Tables S1–S3. Table S1 compiles all time constants extracted from extracavity datasets, Table S2 presents results obtained under resonant VSC conditions, and Table S3 presents results obtained under detuned intracavity conditions. Averaged values and associated standard deviations for each dataset are reported in Table 1 of the main text.

For data processing, some time constants are discarded according to criteria designed to eliminate unreliable values. Any value falling more than three standard deviations from the mean for a given condition is considered an outlier and excluded. In some scans, particular kinetic components cannot be clearly resolved and are likewise discarded. Because we fit each kinetic component independently, exclusion of one component does not affect the inclusion of others in the same dataset. The remaining results demonstrate reproducibility across independent scans, with extracted time constants for extracavity and intracavity conditions all consistent within experimental uncertainty.

**Table S1.** Time constants obtained for the ground-state bleach (GSB) rise and decay and excited-state absorption (ESA) decay and shift from independent UV-pump/IR-probe experiments on extracavity ReCl(CO)$_3$(bpy) in DMSO.

| Extracavity dataset | GSB rise (ps) | GSB decay (ps) | ESA decay (ps) | ESA shift (ps) |
|---|---|---|---|---|
| 2025-06-13-1  | 1.0 | 160 | 100 | 6.5 |
| 2025-06-13-4  | 1.5 | 150 | 120 | 6.2 |
| 2025-06-13-5  | 1.9 | 140 | 130 | 6.4 |
| 2025-06-13-8  | 1.6 | 150 | 140 | 6.5 |
| 2025-06-13-9  | 1.4 | 160 | 120 | 6.5 |
| 2025-06-13-10 | 1.6 | 140 | 170 | 6.5 |
| 2025-06-13-11 | 1.6 | 150 | 90  | 6.4 |
| 2025-06-13-12 | 1.5 | 140 | 140 | 6.8 |
| 2025-06-13-13 | 1.6 | 130 | 100 | 6.4 |
| 2024-06-24-3  | 0.9 |     |     | 6.1 |
| 2024-06-24-4  | 0.9 |     |     | 6.8 |
| 2024-06-24-5  | 1.4 |     |     | 5.8 |
| 2024-06-24-6  | 0.8 |     |     | 6.6 |
| 2024-06-05-2  | 1.2 |     |     |     |
| 2024-06-05-4  | 1.4 |     |     | 4.5 |
| 2024-06-05-5  | 1.1 |     |     | 4.8 |
| 2024-06-06-1  | 1.0 |     |     |     |
| 2024-06-11-2  | 1.7 |     |     | 6.7 |
| 2024-06-11-4  | 1.8 |     |     | 5.4 |
| 2024-06-11-5  | 1.2 |     |     | 5.4 |
| 2024-06-11-6  | 1.3 |     |     | 5.6 |
| 2024-06-11-7  | 1.5 |     |     | 6.5 |
| 2024-06-11-8  | 1.5 |     |     | 5.8 |
| **Mean**      | **1.4 ± 0.3** | **150 ± 10** | **120 ± 20** | **6.1 ± 0.7** |

**Table S2.** Time constants obtained for the ground-state bleach (GSB) rise and decay and excited-state absorption (ESA) decay and shift from independent UV-pump/IR-probe experiments on intracavity ReCl(CO)$_3$(bpy) in DMSO under resonant VSC of the $a'$ symmetric carbonyl stretch at 2018 cm$^{-1}$.

| Resonant VSC dataset | GSB rise (ps) | GSB decay (ps) | ESA decay (ps) | ESA shift (ps) |
|---|---|---|---|---|
| 2025-06-20-1 | 1.0 | 160 | 160 | 4.2 |
| 2025-06-20-2 | 1.1 | 150 | 120 | 4.3 |
| 2025-06-20-3 | 1.7 | 180 | 190 | 4.6 |
| 2025-06-20-4 | 1.5 | 150 | 100 | 4.3 |
| 2025-06-20-5 | 1.3 | 200 | 170 | 5.1 |
| 2025-06-20-6 | 2.0 | 160 | 150 | 4.8 |
| 2025-06-20-7 | 1.2 | 150 | 140 | 3.0 |
| 2025-06-23-1 | 2.9 | 160 | 140 | 4.4 |
| 2024-07-02-1 | 1.5 | | | |
| 2024-07-02-2 | 1.4 | | | |
| 2024-07-02-3 | 1.5 | | | |
| 2024-07-02-4 | 1.2 | | | 7.2 |
| 2024-07-02-5 | 1.4 | | | |
| 2024-07-02-6 | 1.5 | | | 6.1 |
| 2024-07-02-9 | 1.2 | | | 6.5 |
| 2024-07-01-1 | 1.4 | | | |
| 2024-07-01-2 | 1.2 | | | |
| 2024-07-01-3 | 1.5 | | | |
| 2024-07-01-4 | 1.5 | | | |
| 2024-07-01-5 | 1.6 | | | 6.3 |
| 2024-06-27-1 | 1.9 | | | 6.2 |
| 2024-06-27-3 | 2.1 | | | |
| 2024-06-27-5 | 1.8 | | | |
| 2024-06-25-2 | 2.1 | | | |
| **Mean** | **1.6 ± 0.4** | **170 ± 20** | **150 ± 30** | **5.2 ± 1.2** |

**Table S3.** Time constants obtained for the ground-state bleach (GSB) rise and decay and excited-state absorption (ESA) decay and shift from independent UV-pump/IR-probe experiments on intracavity ReCl(CO)$_3$(bpy) in DMSO under detuned intracavity conditions.

| Detuned VSC dataset | GSB rise (ps) | GSB decay (ps) | ESA decay (ps) | ESA shift (ps) |
|---|---|---|---|---|
| 2025-06-23-1 | 1.3 | 150 | 150 | 6.5 |
| 2025-06-23-2 | 1.7 | 160 | 140 | 6.5 |
| 2025-06-23-3 | 1.5 | 180 | 130 | 6.3 |
| 2025-06-24-1 | 2.0 | 130 | 140 | 4.0 |
| 2025-06-24-4 | | 150 | 140 | |
| 2024-07-02-7 | 1.4 | 160 | 90 | 5.6 |
| 2024-06-25-6 | 1.7 | 110 | 100 | 7.1 |
| **Mean** | **1.6 ± 0.3** | **150 ± 20** | **130 ± 30** | **6.0 ± 1.6** |